\documentclass[aps,prc,twocolumn,showpacs,floatfix,nofootinbib]{revtex4}
\usepackage{amsmath,graphicx}

\begin{document}

\title{Relativistic dissipative hydrodynamics from kinetic theory with relaxation time approximation}

\author{Amaresh Jaiswal}
\affiliation{Tata Institute of Fundamental Research,
Homi Bhabha Road, Mumbai 400005, India}

\date{\today}

\begin{abstract}
Starting from Boltzmann equation with relaxation time approximation 
for the collision term and using Chapman-Enskog like expansion for 
distribution function close to equilibrium, we derive hydrodynamic 
evolution equations for the dissipative quantities directly from 
their definition. Although the form of the equations is identical to 
those obtained in traditional Israel-Stewart approaches employing 
Grad's 14-moment approximation and second moment of Boltzmann 
equation, the coefficients obtained are different. In the case of 
one-dimensional scaling expansion, we demonstrate that our results 
are in better agreement with numerical solution of Boltzmann 
equation as compared to Israel-Stewart results. We also show that 
including approximate higher-order corrections in viscous evolution 
significantly improves this agreement, thus justifying the 
relaxation time approximation for the collision term.
\end{abstract}



\maketitle


Relativistic fluid dynamics has been applied quite successfully to 
study and understand a wide range of collective phenomena observed 
in cosmology, astrophysics and the physics of high-energy, heavy-ion 
collisions. The earliest theoretical formulation of relativistic 
dissipative hydrodynamics also known as first-order theories (order 
of gradients), are due to Eckart \cite{Eckart:1940zz} and 
Landau-Lifshitz \cite{Landau}. However these theories, collectively 
called relativistic Navier-Stokes (NS) theory, involve parabolic 
differential equations and suffer from acausality and numerical 
instability. The Chapman-Enskog (CE) expansion has been the most 
common method to obtain first-order hydrodynamics from Boltzmann 
Equation (BE) \cite{Chapman}. The derivation of second-order 
fluid-dynamics by Israel and Stewart (IS) from kinetic theory uses 
extended Grad's method \cite{Israel:1979wp}. The approach by Israel 
and Stewart may not guarantee stability but solves the acausality 
problem \cite{Huovinen:2008te} at the cost of introducing two 
additional approximations: \textbf {(a)} 14-moment approximation for 
the distribution function and, \textbf {(b)} use of second moment of 
BE to obtain evolution equations for dissipative quantities.

Grad's method, originally proposed for non-relativistic systems, was 
modified by Israel and Stewart so that it could be applicable to the 
relativistic case. In this extension, known as 14-moment 
approximation, the distribution function is Taylor expanded in 
powers of four-momenta around its local equilibrium value. 
Truncating the Taylor expansion at second-order in momenta results 
in 14 unknowns that have to be determined to describe the 
distribution function. This expansion implicitly assumes a 
converging series in powers of momenta. In addition, it is assumed 
that the order of expansion in 14-moment approximation (expanded as 
a series in momenta) coincides with that of gradient expansion of 
hydrodynamics. This is evident because Grad's approximation 
truncated at second-order in momenta is not consistent with 
second-order hydrodynamics.

Another assumption inherent in IS derivation is the choice of second 
moment of the BE to extract the equation of motion for the 
dissipative quantities. This choice is arbitrary in the sense that 
once the distribution function is specified, any moment of the BE 
will lead to a closed set of equations for the dissipative currents 
but with different transport coefficients. In fact, it has been 
pointed out in Ref. \cite{Denicol:2010xn} that instead of this 
ambiguous choice of the second-moment of BE by IS, the dissipative 
quantities can be obtained directly from their definition. 
Consistent and accurate formulation of relativistic dissipative 
hydrodynamics is still unresolved and is currently an active 
research area \cite{Denicol:2010xn,Jaiswal:2013fc,Jaiswal:2012qm,El:2009vj,Denicol:2012cn}.

In this Rapid Communication, we present an alternative derivation of 
hydrodynamic equations for dissipative quantities which do not make 
use of both these assumptions. We revisit the CE expansion of the 
distribution function using BE in Relaxation Time Approximation 
(RTA). Using this expansion, we derive the first and second-order 
equations of motion for the dissipative quantities from their 
definition. In one-dimensional boost-invariant Bjorken scenario, we 
demonstrate that our second-order results are in better agreement 
with transport results as compared to those obtained by using IS 
equations. We also illustrate that heuristic incorporation of 
higher-order corrections in viscous evolution equation significantly 
improves this agreement.


Fluid dynamics is best described as a long-wavelength, low-frequency 
limit of an underlying microscopic theory. Further, BE governs the 
temporal evolution of single particle phase-space distribution function 
$f\equiv f(x,p)$ which provides a reliably accurate description of 
the microscopic dynamics in the dilute limit. With this motivation, 
our starting point for the derivation of hydrodynamic equations is 
relativistic BE with RTA for the collision 
term \cite{Anderson_Witting}
\begin{equation}\label{RBE}
p^\mu\partial_\mu f =  -\frac{u\cdot p}{\tau_R}(f-f_0)~,
\end{equation}
where, $p^{\mu}$ is the particle four-momentum, $u_\mu$ is the fluid 
four-velocity and $\tau_R$ is the relaxation time. We define the 
scalar product $u\cdot p\equiv u_\mu p^\mu$.  With $f\to \bar f$ and 
$f_0\to \bar f_0$, Eq. (\ref{RBE}) describes the evolution of 
distribution function for antiparticles. The equilibrium 
distribution functions for Fermi, Bose, and Boltzmann particles 
($r=1,-1,0$) are
\begin{equation}\label{EDF}
f_0=\frac{1}{\mathrm{exp}(\beta\,u\cdot p-\alpha)+r}~,
\end{equation}
and $\alpha\to -\alpha$ for antiparticles $\bar f_0$. Here, 
$\beta=1/T$ is the inverse temperature and $\alpha=\mu/T$ is the 
ratio of chemical potential to temperature.

In the CE expansion, the particle distribution function is expanded 
about its equilibrium value in powers of space-time gradients.
\begin{equation}\label{CEE}
f = f_0 + \delta f, \quad \delta f= \delta f^{(1)} + \delta f^{(2)} + \cdots ,
\end{equation}
where $\delta f^{(1)}$ is first-order in gradients, $\delta f^{(2)}$ 
is second-order and so on. The Boltzmann equation, (\ref{RBE}), in the 
form $f=f_0-(\tau_R/u\cdot p)\,p^\mu\partial_\mu f$, can be solved 
iteratively as \cite{Romatschke:2011qp}
\begin{equation}\label{F1F2}
f_1 = f_0 -\frac{\tau_R}{u\cdot p} \, p^\mu \partial_\mu f_0, \quad f_2 = f_0 -\frac{\tau_R}{u\cdot p} \, p^\mu \partial_\mu f_1, ~~\, \cdots
\end{equation}
where $f_1=f_0+\delta f^{(1)}$ and $f_2=f_0+\delta f^{(1)}+\delta 
f^{(2)}$. To first and second-order in gradients, we obtain
\begin{align}
\delta f^{(1)} &= -\frac{\tau_R}{u\cdot p} \, p^\mu \partial_\mu f_0~, \label{FOC} \\
\delta f^{(2)} &= \frac{\tau_R}{u\cdot p}p^\mu p^\nu\partial_\mu\Big(\frac{\tau_R}{u\cdot p} \partial_\nu f_0\Big)~. \label{SOC}
\end{align}
The above treatment to obtain $\delta f$ is valid for $\delta\bar f$ 
(antiparticles) as well.

For the sake of comparison, we also write down the Grad's 14-moment 
expansion \cite{Grad} in orders of momenta as suggested by IS 
\cite{Israel:1979wp} in orthogonal basis \cite{Denicol:2012cn},
\begin{equation}\label{FME}
\delta f= f_0 \tilde f_0\left( \lambda_\Pi \Pi + \lambda_n n_\alpha p^\alpha 
+ \lambda_\pi \pi_{\alpha\beta} p^\alpha p^\beta \right) + \mathcal{O}(p^3)~,
\end{equation}
where, $\tilde f_0=1-rf_0$ and $\lambda_\Pi$, $\lambda_n$, 
$\lambda_\pi$ are determined from the definition of the dissipative 
quantities, Eqs. (\ref{FBE})-(\ref{FSE}). Since hydrodynamics 
involves expansion in orders of gradients, hence for consistency, CE 
should be preferred over 14-moment approximation in derivation of 
hydrodynamic equations. 


The conserved energy-momentum tensor and particle current can be 
expressed in terms of distribution function as \cite{deGroot}
\begin{align}\label{NTD}
T^{\mu\nu} &= \!\int\! dp \ p^\mu p^\nu (f+\bar f) = \epsilon u^\mu u^\nu-(P+\Pi)\Delta ^{\mu \nu} 
+ \pi^{\mu\nu}~,  \nonumber\\
N^\mu &= \!\int\! dp \ p^\mu (f-\bar f) = nu^\mu + n^\mu~,
\end{align}
where $dp = g d{\bf p}/[(2 \pi)^3\sqrt{{\bf p}^2+m^2}]$, $g$ and $m$ 
being the degeneracy factor and particle mass. In the tensor 
decompositions, $\epsilon, P, n$ are respectively energy density, 
pressure, net number density, and $\Delta^{\mu\nu}=g^{\mu\nu}-u^\mu 
u^\nu$ is the projection operator on the three-space orthogonal to 
the hydrodynamic four-velocity $u^\mu$ defined in the Landau frame: 
$T^{\mu\nu} u_\nu=\epsilon u^\mu$. The metric tensor is 
$g^{\mu\nu}\equiv\mathrm{diag}(+,-,-,-)$. The bulk viscous pressure 
$(\Pi)$, shear stress tensor $(\pi^{\mu\nu})$ and particle diffusion 
current $(n^\mu)$ are the dissipative quantities. 

Energy-momentum conservation, $\partial_\mu T^{\mu\nu} =0$ and 
current conservation, $\partial_\mu N^{\mu}=0$,  yields the 
fundamental evolution equations for $n$, $\epsilon$ and $u^\mu$
\begin{align}\label{evol}
\dot\epsilon + (\epsilon+P+\Pi)\theta - \pi^{\mu\nu}\nabla_{(\mu} u_{\nu)} &= 0~,  \nonumber\\
(\epsilon+P+\Pi)\dot u^\alpha - \nabla^\alpha (P+\Pi) + \Delta^\alpha_\nu \partial_\mu \pi^{\mu\nu}  &= 0~, \nonumber\\
\dot n + n\theta + \partial_\mu n^{\mu} &=0~. 
\end{align}
We use the standard notation $\dot A=u^\mu\partial_\mu A$ for 
co-moving derivative, $\nabla^\alpha=\Delta^{\mu\alpha}\partial_\mu$ 
for space-like derivative, $\theta=\partial_\mu u^\mu$ for expansion 
scalar and $A^{(\alpha}B^{\beta )}=(A^\alpha B^\beta + A^\beta 
B^\alpha)/2$ for symmetrization.

Even if the equation of state relating $\epsilon$ and $P$ is 
provided, the system of Eqs. (\ref{evol}) is not closed unless the 
dissipative quantities $\Pi$, $n^\mu$ and $\pi^{\mu\nu}$ are 
specified. To obtain the expressions for these dissipative 
quantities, we write them using Eq. (\ref{NTD}) in terms of away 
from equilibrium part of the distribution functions ($\delta 
f,~\delta\bar f$) as
\begin{align}
\Pi &= -\frac{\Delta_{\alpha\beta}}{3} \int dp \, p^\alpha p^\beta (\delta f+\delta\bar f) ~, \label{FBE}\\
n^\mu &= \Delta^\mu_\alpha \int dp \, p^\alpha (\delta f-\delta\bar f) ~, \label{FCE}\\
\pi^{\mu\nu} &= \Delta^{\mu\nu}_{\alpha\beta} \int dp \, p^\alpha p^\beta (\delta f+\delta\bar f) ~,\label{FSE}
\end{align}
where
$\Delta^{\mu\nu}_{\alpha\beta} = [\Delta^{\mu}_{\alpha}\Delta^{\nu}_{\beta} + \Delta^{\mu}_{\beta}\Delta^{\nu}_{\alpha} - (2/3)\Delta^{\mu\nu}\Delta_{\alpha\beta}]/2$.

The first-order dissipative equations can be obtained from Eqs. 
(\ref{FBE})-(\ref{FSE}) using $\delta f = \delta f^{(1)}$ from Eq. 
(\ref{FOC})
\begin{align}
\Pi &= -\frac{\Delta_{\alpha\beta}}{3}\!\!\int\!\! dp \, p^\alpha p^\beta \left[-\frac{\tau_R}{u.p} \, p^\gamma \partial_\gamma \left(f_0+\bar f_0\right)\right] , \label{FOBE}\\
n^\mu &= \Delta^\mu_\alpha \!\!\int\!\! dp \, p^\alpha \left[-\frac{\tau_R}{u.p} \, p^\gamma \partial_\gamma \left(f_0-\bar f_0\right)\right] , \label{FOCE}\\
\pi^{\mu\nu} &= \Delta^{\mu\nu}_{\alpha\beta}\int dp \ p^\alpha p^\beta \left[-\frac{\tau_R}{u.p} \, p^\gamma \partial_\gamma \left(f_0+\bar f_0\right)\right] . \label{FOSE}
\end{align}
Assuming the relaxation time $\tau_R$ to be independent of 
four-momenta, the integrals in Eqs. (\ref {FOBE})-(\ref{FOSE}) 
reduce to
\begin{equation}\label{FOE}
\Pi = -\tau_R\beta_\Pi\theta, ~~  n^\mu = \tau_R\beta_n\nabla^\mu\alpha, ~~ \pi^{\mu\nu} = 2\tau_R\beta_\pi\sigma^{\mu\nu},
\end{equation}
where 
$\sigma^{\mu\nu}=\Delta^{\mu\nu}_{\alpha\beta}\nabla^{\alpha}u^\beta$. 
The coefficients $\beta_\Pi,~\beta_n$ and $\beta_\pi$ are 
found to be
\begin{align}
\beta_\Pi =&~ \frac{1}{3}\left(1-3c_s^2\right)(\epsilon+P) - \frac{2}{9}(\epsilon-3P)\nonumber\\ 
&- \frac{m^4}{9}\left<(u.p)^{-2}\right>_{0^+} , \label{BB}\\
\beta_n =&~ - \frac{n^2}{\beta(\epsilon+P)} + \frac{2\left<1\right>_{0^-}}{3\beta} + \frac{m^2}{3\beta}\left<(u.p)^{-2}\right>_{0^-} ,\label{BC}\\
\beta_\pi =&~ \frac{4P}{5} + \frac{\epsilon-3P}{15} - \frac{m^4}{15}\left<(u.p)^{-2}\right>_{0^+} ,\label{BS}
\end{align}
where $\left<\cdots\right>_{0^\pm}=\int dp(\cdots)(f_0\pm\bar f_0)$, 
and $c_s^2=(dP/d\epsilon)_{s/n}$ is the adiabatic speed of sound squared 
($s$ being the entropy density). It is interesting to note that these 
coefficients are in perfect agreement with those obtained in the 
Ref. \cite {Denicol:2010xn} in which the evolution equations are 
derived directly from their definition. This is due to the fact that 
in Ref. \cite{Denicol:2010xn}, the coefficients $\beta_\Pi,~\beta_n$ 
and $\beta_\pi$, are associated with first-order terms and do not 
involve 14-moment approximation. In the massless limit, 
$\beta_\pi=4P/5$ is also in agreement with that obtained in Ref. 
\cite {Romatschke:2011qp} employing CE expansion in BE with 
medium-dependent masses.

In the process to obtain second-order equations, we discover that CE 
expansion for the distribution function does not support derivation 
of hydrodynamic evolution equations from arbitrary moment choice of 
BE. Using the definition of dissipative quantities to obtain their 
evolution equations comes naturally when employing CE expansion as 
demonstrated while deriving first-order equations, Eq. (\ref{FOE}). 
Second-order evolution equations can also be obtained similarly by 
substituting $\delta f=\delta f^{(1)}+\delta f^{(2)}$ from Eqs. (\ref
{FOC}) and (\ref{SOC}) in Eqs. (\ref {FBE})-(\ref {FSE}).
\begin{align} 
\frac{\Pi}{\tau_R} =& \
\frac{\Delta_{\alpha\beta}}{3}\!\!\int\!\! dp \, p^\alpha p^\beta \Big[\frac{p^\gamma}{u\cdot p}\partial_\gamma f_0 
- \frac{p^\gamma p^\rho}{u\cdot p}\partial_\gamma\Big(\frac{\tau_R}{u\cdot p}\partial_\rho f_0\Big)\nonumber \\
&\quad\quad\quad\quad\quad\quad\quad +f_0\to\bar f_0\Big] ~, \label{SOBE}\\
\frac{n^\mu}{\tau_R} =& 
-\Delta^\mu_\alpha \!\!\int\!\! dp \, p^\alpha \Big[\frac{p^\gamma}{u\cdot p}\partial_\gamma f_0 
- \frac{p^\gamma p^\rho}{u\cdot p}\partial_\gamma\Big(\frac{\tau_R}{u\cdot p}\partial_\rho f_0\Big)\nonumber \\
&\quad\quad\quad\quad\quad\quad -f_0\to\bar f_0\Big] ~, \label{SOCE}\\
\frac{\pi^{\mu\nu}}{\tau_R} =& 
-\Delta^{\mu\nu}_{\alpha\beta}\!\!\int\!\! dp \, p^\alpha p^\beta \Big[\frac{p^\gamma}{u\cdot p}\partial_\gamma f_0 
- \frac{p^\gamma p^\rho}{u\cdot p}\partial_\gamma\Big(\frac{\tau_R}{u\cdot p}\partial_\rho f_0\Big)\nonumber \\
&\quad\quad\quad\quad\quad\quad\quad\quad +f_0\to\bar f_0\Big] ~. \label{SOSE}
\end{align}

The derivatives of equilibrium distribution function ($\partial_\mu 
f_0,~\partial_\mu \partial_\nu f_0$) appearing in above equations 
can be obtained by successively differentiating Eq. (\ref{EDF}). The 
momentum integrations can be decomposed into hydrodynamic tensor 
degrees of freedom via the definitions:
\begin{align}\label{TDI}
I^{\mu_1\cdots\mu_n}_{(m)\pm} \!\equiv \!\!&\int\!\!\! \frac{dp}{(u\!\cdot\! p)^m} p^{\mu_1}\!\cdots p^{\mu_n} (f_0 \pm \bar f_0) 
\!=\! I_{n0}^{(m)\pm} u^{\mu_1}\!\cdots u^{\mu_n}  \nonumber \\
&+ I_{n1}^{(m)\pm} (\Delta^{\mu_1\mu_2} u^{\mu_3} \cdots u^{\mu_n} + \mathrm{perms}) + \cdots,
\end{align}
where `perms' denotes all non-trivial permutations of the Lorentz 
indices. We similarly define $J^{\mu_1\mu_2\cdots\mu_n}_{(m)\pm}$ 
where the momentum integrals are weighted with $f_0 \tilde f_0 \pm 
(f_0 \to \bar f_0)$, and are tensor decomposed with coefficients 
$J_{nq}^{(m)\pm}$. 

After performing the integration, the relaxation time appearing on 
the right hand side of Eqs. (\ref {SOBE})-(\ref{SOSE}) are absorbed 
using the first-order equations for the dissipative quantities, Eq. 
(\ref{FOE}). Using the identity $\nabla^\mu\beta=-\beta\dot 
u^\mu+[n/(\epsilon+P)]\nabla^\mu\alpha+\mathcal{O}(\delta^2)$, the 
terms containing derivatives of the relaxation time cancel each 
other upto second-order in gradients and hence the right hand side 
of Eqs. (\ref {SOBE})-(\ref {SOSE}) can be made independent of 
$\tau_R$ \cite {supplement}. The second-order evolution equations of 
the dissipative quantities are finally obtained as
\begin{align}
\frac{\Pi}{\tau_R} =& -\dot{\Pi}
-\beta_{\Pi}\theta 
-\delta_{\Pi\Pi}\Pi\theta
+\lambda_{\Pi\pi}\pi^{\mu\nu}\sigma_{\mu \nu }  \nonumber \\
&-\tau_{\Pi n}n\cdot\dot{u}
-\lambda_{\Pi n}n\cdot\nabla\alpha
-\ell_{\Pi n}\partial\cdot n ~, \label{BULK}\\
\frac{n^{\mu}}{\tau_R} =& -\dot{n}^{\langle\mu\rangle}
+\beta_{n}\nabla^{\mu}\alpha
-n_{\nu}\omega^{\nu\mu}
-\lambda_{nn}n^{\nu}\sigma_{\nu}^{\mu}
-\delta_{nn}n^{\mu}\theta   \nonumber \\
&+\lambda_{n\Pi}\Pi\nabla^{\mu}\alpha
-\lambda_{n\pi}\pi^{\mu\nu}\nabla_{\nu}\alpha 
-\tau_{n\pi}\pi_{\nu}^{\mu}\dot{u}^{\nu}  \nonumber \\
&+\tau_{n\Pi}\Pi\dot{u}^{\mu}
+\ell_{n\pi}\Delta^{\mu\nu}\partial_{\gamma}\pi_{\nu}^{\gamma}
-\ell_{n\Pi}\nabla^{\mu}\Pi~,  \label{HEAT} \\
\frac{\pi^{\mu\nu}}{\tau_R} =& -\dot{\pi}^{\langle\mu\nu\rangle}
+2\beta_{\pi}\sigma^{\mu\nu}
+2\pi_{\gamma}^{\langle\mu}\omega^{\nu\rangle\gamma}
-\tau_{\pi\pi}\pi_{\gamma}^{\langle\mu}\sigma^{\nu\rangle\gamma}  \nonumber \\
&-\delta_{\pi\pi}\pi^{\mu\nu}\theta 
+\lambda_{\pi\Pi}\Pi\sigma^{\mu\nu}
-\tau_{\pi n}n^{\langle\mu}\dot{u}^{\nu\rangle }  \nonumber \\
&+\lambda_{\pi n}n^{\langle\mu}\nabla ^{\nu\rangle}\alpha
+\ell_{\pi n}\nabla^{\langle\mu}n^{\nu\rangle } ~, \label{SHEAR}
\end{align}
where the vorticity tensor is defined as 
$\omega^{\mu\nu}=(\nabla^\mu u^\nu-\nabla^\nu u^\mu)/2$. All the 
coefficients in the above equations have been obtained in terms of 
$\beta$ and the integral coefficients $I_{nq}^{(m)\pm}$ and 
$J_{nq}^{(m)\pm}$ \cite {supplement}. It is clear that in Eqs. (\ref 
{BULK})-(\ref {SHEAR}), the Boltzmann relaxation time $\tau_R$ can 
be replaced by those of the individual dissipative quantities 
$\tau_\Pi,~\tau_n,~\tau_\pi$. At this stage, it seems as though the 
three relaxation times $\tau_\Pi,~\tau_n,~\tau_\pi$ are all equal to 
$\tau_R$. This is because the collision term in the BE, Eq. (\ref 
{RBE}) is written in RTA which does not entirely capture the 
microscopic interactions. This apparent equality vanishes if the 
first-order equation, Eq. (\ref {FOE}) is compared with the 
relativistic Navier-Stokes equations for dissipative quantities 
($\Pi=-\zeta\theta$, $n^\mu=\lambda T\nabla^\mu \alpha$ and 
$\pi^{\mu\nu}=2\eta\sigma^{\mu\nu}$). The dissipative relaxation 
times are then obtained in terms of first-order transport 
coefficients $\zeta,~\lambda$ and $\eta$ which can be calculated 
independently taking into account the full microscopic behavior of 
the system \cite {Kovtun:2004de,Meyer:2007dy}. 

We remark that although the form of the evolution equations for 
dissipative quantities obtained here, Eqs. (\ref{BULK})-(\ref 
{SHEAR}), are the same as those obtained in the previous 
calculations using both 14-moment approximation and second moment of 
BE \cite{Betz:2008me}, the coefficients obtained are different. In 
the following discussion, we refer to the results in Ref. \cite
{Betz:2008me} as the IS results although the power counting scheme 
differs from the one employed originally by IS. 

For the special case of a system consisting of single species of 
massless Boltzmann gas, we find that
\begin{equation}\label{TDCE}
\beta_\pi = \frac{4P}{5}, \quad \tau_{\pi\pi} = \frac{10}{7}, \quad \delta_{\pi\pi} = \frac{4}{3};
\end{equation}
while these coefficients obtained via IS approach are 
\cite{Betz:2008me}
\begin{equation}\label{TDIS}
\beta_\pi^{\rm{IS}} = \frac{2P}{3}, \quad \tau_{\pi\pi}^{\rm{IS}} = 2, \quad \delta_{\pi\pi}^{\rm{IS}} = \frac{4}{3}.
\end{equation}
In this limit, although the coefficients of $\pi^{\mu\nu}\theta$ are 
same for both the cases ($\delta_{\pi\pi}=\delta_{\pi\pi}^{\rm{IS}}$), 
the coefficient of $\sigma^{\mu\nu}$ and 
$\pi_{\gamma}^{\langle\mu}\sigma^{\nu\rangle\gamma}$ are different 
($\beta_\pi\ne\beta_\pi^{\rm{IS}},~\tau_{\pi\pi}\ne\tau_{\pi\pi}^{\rm{I
S}}$).

We note that CE expansion, as opposed to 14-moment approximation, 
can be done iteratively to arbitrarily higher orders. Hence using CE 
expansion, dissipative hydrodynamic equations of any order can in 
principle be derived. To obtain $n$th-order evolution equations for 
dissipative quantities, $\delta f=\delta f^{(1)}+\delta 
f^{(2)}+\cdots+\delta f^{(n)}$ should be used in Eqs. (\ref{FBE})-
(\ref {FSE}). For instance, substitution of $\delta f=\delta 
f^{(1)}+\delta f^{(2)}+\delta f^{(3)}$ in Eqs. (\ref{FBE})-(\ref 
{FSE}) will eventually lead to third-order evolution equations. 
Derivation of third-order hydrodynamics as outlined above is left 
for future work.


To demonstrate the numerical significance of the new coefficients 
derived here, we consider evolution in the boost invariant Bjorken 
case of a massless Boltzmann gas ($\epsilon=3P$) at vanishing net 
baryon number density \cite {Bjorken:1982qr}. In terms of the Milne 
co-ordinates $(\tau,x,y,\eta)$, where $\tau = \sqrt{t^2-z^2}$ and 
$\eta=\tanh^{-1}(z/t)$, the initial four-velocity becomes 
$u^\mu=(1,0,0,0)$. For this scenario, $\Pi=n^\mu=0$, and the 
evolution equations for $\epsilon$, $\pi\equiv-\tau^2\pi^{\eta\eta}$ 
reduces to
\begin{align}
\frac{d\epsilon}{d\tau} &= -\frac{1}{\tau}\left(\epsilon + P  -\pi\right), \label{BED} \\
\frac{d\pi}{d\tau} &= - \frac{\pi}{\tau_R} + \beta_\pi\frac{4}{3\tau} - \lambda\frac{\pi}{\tau}. \label{Bshear}
\end{align}
The second-order transport coefficients simplify to
\begin{equation}\label{BTC}
\lambda \equiv \frac{1}{3}\tau_{\pi\pi}+\delta_{\pi\pi} = \frac{38}{21}, \quad \lambda^{\rm{IS}}=2.
\end{equation}

\begin{figure}[t]
\begin{center}
\includegraphics[scale=0.5]{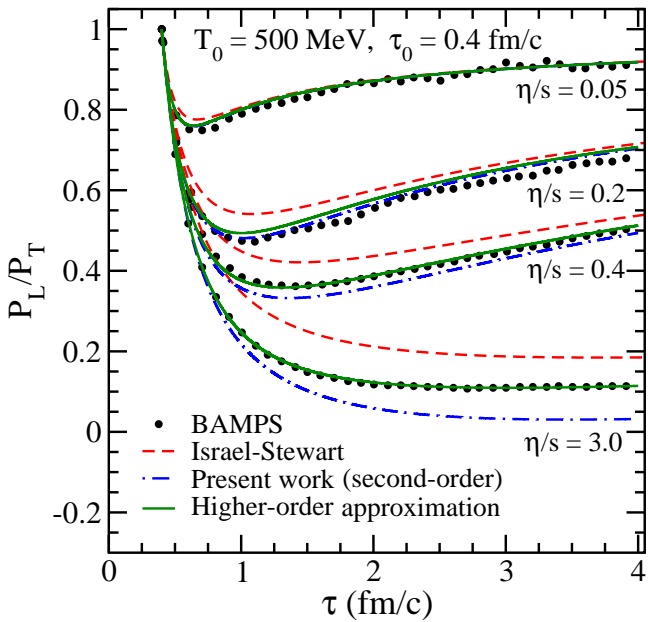}
\end{center}
\vspace{-0.4cm}
\caption{(Color online) Time evolution of $P_L/P_T$ in BAMPS (dots), IS (dashed
  lines), present work (dashed-dotted line), and a heuristic 
  higher-order approximation (solid line) for isotropic initial 
  pressure configuration $(\pi_0=0)$.}
\label{PLPT}
\end{figure}
Initial temperature $T_0=500$ MeV at proper time $\tau_0=0.4$ fm/c 
are chosen to solve the coupled differential Eqs. (\ref{BED}) and 
(\ref {Bshear}). These values correspond to LHC initial conditions 
\cite{El:2007vg}. We assume isotropic initial pressure configuration 
i.e. $\pi_0=0$. Fig. \ref{PLPT}, shows the proper time dependence of 
pressure anisotropy defined as $P_L/P_T= (P-\pi)/(P+\pi/2)$. The 
dashed and dashed-dotted lines represent the results from IS theory 
and our second-order results, respectively. The dots correspond to 
the results of a transport model, the Boltzmann Approach of 
MultiParton Scatterings (BAMPS), which is based on parton cascade 
simulations \cite {El:2009vj,Xu:2004mz}. The calculations in BAMPS 
are performed with variable values for the cross section such that 
the shear viscosity to entropy density ratio is a constant.

We note that the results from IS theory always overestimate the 
pressure anisotropy as compared to the transport results even for 
viscosities as small as $\eta/s=0.05$. It is evident from the figure 
that our results are in better agreement with BAMPS as compared to 
the results of IS. For very high viscosity, i.e., for $\eta/s=3.0$, 
although at early times we have a better agreement with BAMPS as 
compared to IS, at later times there is a large deviation. This 
disagreement may be attributed to the fact that the present 
hydrodynamic calculation is terminated at second-order in gradients. 
Inclusion of higher-order corrections may improve the agreement of 
dissipative hydrodynamic calculation results with those obtained 
using BAMPS as illustrated in the following.

In Ref. \cite{El:2009vj}, while performing a third-order calculation 
it was demonstrated that within one-dimensional scaling expansion, 
the higher-order gradient terms can acquire the form 
$(\frac{\pi}{\epsilon})^n\frac{\epsilon}{\tau}$, where, $n=r-1$ for 
$r$th-order corrections. The other forms of higher-order corrections 
is reducible to this structure through lower-order evolution 
equations. Here we assume a similar heuristic expression for 
higher-order corrections
\begin{equation}
\frac{d\pi}{d\tau} = - \frac{\pi}{\tau_R} + \beta_\pi\frac{4}{3\tau} 
- \lambda\frac{\pi}{\tau} - \chi\frac{\pi^2}{\beta_\pi\tau}, \label{HOA}
\end{equation}
where the coefficient $\chi$ contains corrections to shear stress 
evolution due to higher-order gradients. This coefficient can be 
obtained by demanding that the above equation be valid for a free 
streaming of particles in the limit of infinite shear viscosity 
($\eta\to\infty$). In this limit, $\tau_R\to\infty$, and within 
one-dimensional scaling expansion the energy density evolves as 
$\dot\epsilon=-\epsilon/\tau$ which implies that $\dot P=-P/\tau$. 
For this case, using Eq. (\ref{BED}), we arrive at $\pi=\epsilon/3=P$
which indicates disappearance of the longitudinal pressure. 
Substituting all these in Eq. (\ref{HOA}), we obtain $\chi=36/175$.

Fig. \ref{PLPT}, also shows $P_L/P_T$ evolution for the results 
obtained after including higher-order corrections (solid lines). We 
observe that the incorporation of higher-order corrections 
significantly improves the agreement with BAMPS. It is important to 
note that the BAMPS calculations are performed with the form of the 
collision term that captures the realistic microscopic interactions 
whereas the derivation of dissipative hydrodynamic equations in the 
present work uses RTA for the collision term. Within CE formalism, 
more sophisticated ways exist for solving the BE, for eg., by using 
variational methods \cite{Chapman} or by considering momentum 
dependent relaxation time \cite {Prakash:1993bt,Teaney:2009qa}. It 
is, in principle, possible to derive second-order dissipative 
hydrodynamic evolution equations using momentum dependent relaxation 
time provided the dependence is specified explicitly. While this is 
left for future work, we observe that the near perfect agreement of 
the BAMPS results with those obtained using higher-order corrections 
clearly suggest that the momentum independent relaxation time for 
the BE used in the present derivation is sufficiently reliable for 
the range of $\eta/s$ considered here. However, the results obtained 
by using a momentum dependent relaxation time may show a better 
agreement with BAMPS data already at second-order.

RTA for the collision term assumes that the effect of the collisions 
is to restore the distribution function to its local equilibrium 
value exponentially. This is a very good approximation as long as 
the deviations from local equilibrium are small. As discussed above, 
we find that for the range of $\eta/s$ considered here, the 
deviation from equilibrium is not so large because the RTA is still 
valid. It is also important to note that large values of $\eta/s$ 
($> 0.4$) are not relevant to the physics of strongly coupled systems 
like Quark Gluon Plasma (QGP). The QGP formed at RHIC and LHC 
behaves as a near perfect fluid with a small estimated 
$\eta/s\approx 0.08-0.2$ \cite {Romatschke:2007mq,Song:2010mg}. 
Using second-order evolution equations derived here, we get 
reasonably good agreement with BAMPS results for $\eta/s\le 0.4$ 
(Fig. \ref {PLPT}). This suggests that BE with RTA for the collision 
term can be successfully applied in understanding the hydrodynamic 
behaviour of QGP formed in relativistic heavy-ion collisions. 

To summarize, we have presented a new derivation of relativistic 
second-order hydrodynamics from BE. We use Chapman-Enskog expansion 
for out of equilibrium distribution function instead of 14-moment 
approximation and derive evolution equations for dissipative 
quantities directly from their definitions rather than employing 
second moment of Boltzmann equation. In this new approach, we get 
rid of two powerful assumptions of Israel-Stewart kind of derivation 
which is 14-moment approximation and choice of second moment of 
Boltzmann equation. Although the form of the evolution equation 
remains the same, the coefficients are found to be different. For 
small $\eta/s$, our second-order results show reasonably good 
agreement with the parton cascade BAMPS for the $P_L/P_T$ evolution. 
We find that heuristic inclusion of higher-order corrections in 
shear evolution equation significantly improves the agreement with 
transport calculation for large $\eta/s$ as well. This concurrence 
also suggests that relaxation time approximation for the collision 
term in Boltzmann equation is reasonably accurate when applied to 
heavy-ion collisions.

\begin{acknowledgments}
The author would like to thank Rajeev S. Bhalerao and Subrata Pal 
for fruitful discussions, critical reading of the manuscript and 
helpful comments.
\end{acknowledgments}

\end{document}